\documentclass[aps,prl,twocolumn,showpacs,a4paper]{revtex4}
\usepackage{graphicx}

\begin{document} 
\title{ Mechanical measurement of equilibrium spin currents in the Rashba medium}  
\author{ E.B. Sonin}

\affiliation{ Racah Institute of Physics, Hebrew University of
Jerusalem, Jerusalem 91904, Israel} 

\date{\today} 

\begin{abstract}
We demonstrate that an equilibrium spin current in a 2D electron gas with Rashba spin-orbit interaction (Rashba medium) results in a mechanical  torque on a substrate  near an edge of the medium. If the substrate is a cantilever, the mechanical torque displaces the free end of the cantilever. The effect can be  enhanced and tuned  by a magnetic field. Observation of this displacement would be  an effective method to prove existence of equilibrium spin currents. The analysis of edges of the Rashba medium demonstrates the existence of localized edge states.  They form a 1D continuum of states. This suggests a new type of quantum wire: spin-orbit quantum wire.
\end{abstract} 


\maketitle

There is a long-standing interest to the problem of dissipationless spin transport in various media. In the past the focus was on magnetically ordered systems and superfluid $^3$He phases \cite{Usp,Bunk}. Nowadays, spin transport is an essential part of spintronics \cite{ZFS}. A system of special interest is the 2D electron gas with the Rashba spin-orbit interaction (Rashba medium), where they hope to realize effective methods to manipulate spins with an electric field. On the other hand, the concept of {\em spin current} itself arose a lot of questions, in particular, those connected with the absence of the total-spin conservation.  Especially intriguing  seemed that in the Rashba medium  spin currents appear even at the equilibrium, though they do not result in any  accumulation of spin \cite{R}.  It was put in question whether equilibrium spin currents have any relation to real spin transport, and the  presence of spin currents in the ground state was considered as an inherent problem in the spin current concept  \cite{R1}.

In Ref. \onlinecite{BR} it was shown that in the Rashba medium with a spatially modulated spin-orbit parameter there are areas where spin is created or absorbed, and spin currents transport spin from  areas of creation to areas of absorption. This pointed out that at least some relation of equilibrium spin currents to spin transport should not be ruled out.  But the central question for understanding the physical sense of the spin current is  {\em how is it possible, if possible et all, to  detect the existence of spin currents experimentally}.  The present Letter attracts attention  to the fact that equilibrium spin currents in the Rashba medium must lead to mechanical torques on a substrate at edges of the Rashba medium. 
If the substrate is flexible, the torques should deform it,  and measurement of  this deformation would provide a method to detect equilibrium spin currents experimentally.  An appropriate   experimental technique for such a measurement is already known: One may use a {\em mechanical cantilever magnetometer with an integrated 2D electron system} \cite{Gamb}. 

\begin{figure}
\begin{center}
   \leavevmode
  \includegraphics[width=0.9\linewidth]{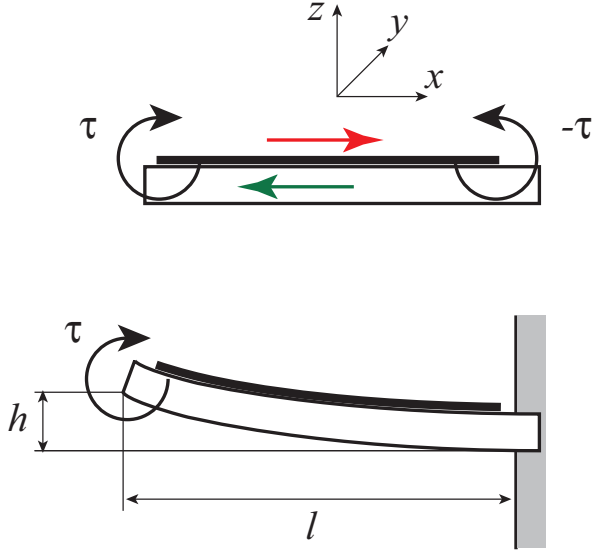}
 \caption{(Color online) The cantilever with  the integrated Rashba medium (thick solid line) on it. (a) A rigid substrate. The red arrow (above the Rashba medium) shows the direction of the spin current. The green arrow (inside the substrate) show the direction of the orbital-moment current. The currents result in mechanical torques $\pm \tau$ at the edges of the substrate (b) The substrate is now a flexible cantilever. The torque at its free end leads to its displacement $h$. }
 \label{fig1}
 \end{center}
\end{figure}

It is possible to explain the appearance of the mechanical toque due to bulk spin currents using simple qualitative arguments based on the conservation laws. Though the Rashba spin-orbit interaction violates the law of spin conservation, it certainly must not violate the conservation law of the total angular momentum which includes the spin and the orbital parts. The continuity equations  for spin and orbital moment  are:
\begin{eqnarray}
{\partial S_\beta \over \partial t}+\nabla_\gamma J_\gamma^\beta  =G_\beta  ~,
          \label{SB}       \end{eqnarray} 
\begin{eqnarray}
{\partial L_\beta \over \partial t}+\nabla_\gamma \tilde J_\gamma^\beta  =-G_\beta  ~.
                \end{eqnarray} 
Here $S_\beta$ and $L_\beta$ are $\beta$ ($\beta,\,\gamma=x,y,z$) components of the spin and the orbital-moment densities, $J_\gamma^\beta$ and $\tilde J_\gamma^\beta$ are corresponding flux-tensors (currents). The equations contain the source terms (torques,) $G_\beta$ and $-G_\beta$, which cancel in the equation for the total angular momentum $\vec S+\vec L$.  Now let us apply these balances to the Rashba medium on a  flexible substrate.  At the equilibrium the time derivatives of the angular momenta are absent, but  there are spin currents inside the Rashba medium \cite{R,BR}.  We consider the current of the $y$ spin component along the  axis $x$, so $\beta=y$ and $\gamma=x$ (Fig. \ref{fig1}). 
In the bulk  there is no torque $\vec G$, and the spin current is constant. 
On the other hand, at the very edge of the Rashba medium the spin current must vanish. These two statements are compatible only if one takes into account non-conservation of spin: there must be  spin torque near the Rashba-medium edge. It compensates the constant bulk spin current so that  $J_x^y  =\int G_y(x) dx$,  where the integral is over the whole edge area of nonzero torque. But the balance equations above require that an  orbital torque and correspondingly   a flux of the orbital moment of opposite sign must appear. Since the 2D electron gas has no orbital moment in its plane, the whole orbital torque must be applied to an edge of the  substrate. Now if the substrate is a cantilever rigidly fixed at one end (see Fig. \ref{fig1}), the mechanical torque $\tau =\tilde J_x^y=-J_x^y =- \int G_y(x) dx$ will deform the cantilever, and the displacement of its free end can be measured.

It is worthwhile to stress that this scenario is not sensitive to what type of edges the Rashba medium has. But in order to demonstrate how this scenario can be realized we analyze below the case of an ideal reflecting boundary of the Rashba medium and show how the spin torque appears near the edges. A side product of this analysis is the prediction that there exist localized electron states near the edges (edge states). The edge states have one-dimensional spectrum forming a new type of the 1D quantum wire: {\em spin-orbit quantum wire}.

The 2D electron gas with Rashba spin-obit interaction is described by the  single-electron hamiltonian
\begin{eqnarray}
H ={\hbar^2\over 2 m}\left\{\vec \nabla \mathbf{\Psi}^\dagger \vec \nabla \mathbf{\Psi} 
 +i\alpha(\vec r) (\mathbf{\Psi}^\dagger  [\vec \sigma \times \hat z]_i \vec\nabla_i\mathbf{\Psi} 
 \right. \nonumber \\ \left.
 -\vec \nabla_i \mathbf{\Psi}^\dagger[\vec \sigma \times \hat z]_i\mathbf{\Psi} )\right\}+V(\vec r) |\mathbf{\Psi}|^2\, ,
     \label{Ham}         \end{eqnarray}
where  $\mathbf{\Psi} =\left( \begin{array}{c} \psi_\uparrow \\ \psi_\downarrow\end{array}\right)$
is a two-component spinor, $\vec \sigma$ is the vector of Pauli matrices, and $V(\vec r)$ is the potential.
In general the spin-orbit parameter $\alpha(\vec r)$ depends on the 2D position vector $\vec r$. The Schr\"odinger equation is:
 \begin{eqnarray}
i\hbar \dot \mathbf{\Psi} 
 ={\hbar^2 \over m}\left\{-{1 \over 2}\nabla^2 \mathbf{\Psi}
 \right. \nonumber \\ \left. 
+i\alpha(\vec r)  [\vec \sigma \times \hat z]_i \nabla_i\mathbf{\Psi}
+{i\over 2} \nabla_i\alpha(\vec r)  [\vec \sigma \times \hat z]_i \mathbf{\Psi}\right\}
+V(\vec r)\mathbf{\Psi}\,.
           \end{eqnarray}           

The expressions for spin current and spin torque in the spin continuity equation (\ref{SB})  can be derived from the Schr\"odinger equation multiplying it by complex-conjugated components: 
\begin{eqnarray}
J_i^\beta=-{i\hbar^2 \over 4 m}( \mathbf{\Psi}^\dagger \sigma_\beta \nabla_i \mathbf{\Psi} -\nabla_i\mathbf{\Psi}^\dagger \sigma_\beta  \mathbf{\Psi}) 
+{\alpha\hbar^2\over 2m} |\mathbf{\Psi}|^2\varepsilon_{z\beta i}\, ,
    \label{SC}         \end{eqnarray}
 \begin{eqnarray}
G_\beta=-  {i\alpha\hbar^2\over 2m}\left\{  \left(\mathbf{\Psi}^\dagger\{ [\vec \sigma \times [\hat z \times \vec  \nabla]]_\beta \mathbf{\Psi}\} \right)
\right. \nonumber \\ \left.
 -\left(\{ [[\vec\nabla \times \hat z]\times \vec \sigma]_\beta \mathbf{\Psi}^\dagger\}   \mathbf{\Psi} \right)\right\}\,.
   \label{torqG}      \end{eqnarray}            
Whereas Greek super(sub)scripts $\beta$ referred to three components $x,y,z$ in the 3D spin space, the Latin super(sub)script $i$ is related with the two coordinates $x,y$ in the 2D electron layer.

Suppose that the Rashba medium occupies the semispace $x<0$ while at $x>0$ spin-orbit interaction is absent. Two semispaces have also different potentials. Thus 
 \begin{eqnarray}
V(\vec r) =\left\{ \begin{array}{cc} 0 &  \mbox{at}~x<0 \\ U  & \mbox{at}~x>0 \end{array}\right. ~,~~
\alpha(\vec r) =\left\{ \begin{array}{cc} \alpha &  \mbox{at}~x<0 \\ 0  & \mbox{at}~x>0 \end{array}\right. ~.
              \end{eqnarray}
The electron states near the interface between two regions with different spin-orbit  constants  have already been analyzed  by Khodas {\em et al} \cite{Fink}. But in contrast to them, in our case the potential barrier $U$ is so high  that electrons cannot  propagate in the area $x>0$. 

The plane-wave solutions of the Schr\"odinger equation are
${1\over \sqrt{2}}\left( \begin{array}{c} 1 \\ \kappa \end{array}\right)e^{i\vec k  \vec r}$, 
where $\kappa = \pm(k_y-ik_x) /k$, and the upper (lower) sign corresponds to the upper (lower) branch of the spectrum (band) with the energies   
 \begin{eqnarray}
\epsilon={\hbar^2(k_0^2-\alpha ^2)\over 2 m}  ={\hbar^2\over m}\left({k^2\over 2}\pm \alpha k\right)~.
           \label{ener}  \end{eqnarray}   
The energy is parametrized by the wave number  $k_0$, which is connected with absolute values of wave vectors in two bands as $k=|k_0 \mp \alpha|$.  
Spin torque in the plane-wave eigenstate is absent, but  there are flows of spin components in the layer plane given by  
 \begin{eqnarray}
J^i_{j\pm}(\vec k) ={\hbar^2 \over 2m}\left(\pm {\varepsilon_{zis} k_s \over k}k_j +\alpha \varepsilon_{zij}\right)~.
             \end{eqnarray}

At $x<0$ one should look for a superposition of plane waves, 
one incident wave, which is coming from $x =-\infty$, and two reflected waves:
\begin{eqnarray}
\psi_\uparrow=e^{ik_y y}\left[\left( \begin{array}{c} 1\\ \kappa_+\end{array}\right)e^{ik_+ x} +r_1\left( \begin{array}{c} 1\\ \bar \kappa_+\end{array}\right)e^{-ik_+ x}
\right. \nonumber \\ \left.
+r_2\left( \begin{array}{c} 1\\ \kappa_-\end{array}\right) e^{ik_- x}) \right]\, ,  
     \label{x<}         \end{eqnarray} 
where $k_\pm =\sqrt{(\alpha \pm k_0)^2-k_y^2}$,  $ \kappa_\pm=-  (k_y-ik_\pm)/k$,  and $ \bar \kappa_+=-  (k_y+ik_+)/k$  are the $x$ components of the wave vectors corresponding to the same energy. We shall consider the case of $k_0<\alpha$, when the both waves belong to the two parts of the lower band to the right ($k>\alpha$) or to the left ($k<\alpha$) from the energy minimum.  The choice of the positive sign before $k_-$ in the exponent of the second reflected wave is determined by the negative group velocity of this wave. Since the electron transport is determined by the group velocity,  the latter should be directed from the boundary inside the bulk even though the wave vector is directed to the boundary. 
At $x>0$ the wave function is evanescent: $\mathbf{\Psi}=\left( \begin{array}{c} t_\uparrow \\ t_\downarrow\end{array}\right)
e^{ik_y y}  e^{-k_b x}$,  where $k_b =\sqrt{2mU/\hbar^2 -k_0^2+\alpha^2}$. 

The wave superposition should satisfy the boundary conditions, which include continuity of the both components of the spinor and jumps of the first derivatives of these components \cite{Fink}: 
 \begin{eqnarray}
\left.{\partial \mathbf{\Psi} \over \partial x }\right |_{+0}-\left.{\partial \mathbf{\Psi}\over \partial x }\right |_{-0}=-  i\alpha(x)  \sigma_ y \mathbf{\Psi} ~.
              \end{eqnarray}

The expressions for the reflection coefficients are rather cumbersome in general, and we present below only the expression for the determinant  of the system of equations (boundary conditions) for the reflection coefficients:
\begin{eqnarray}
D= (\kappa_-- \bar \kappa_+) [k_b^2 -ik_b(k_+-k_-) +\alpha^2+k_+k_-]
\nonumber \\
 -i\alpha(\kappa_- \bar \kappa_++1)(k_++k_-)
~.
            \label{det}       \end{eqnarray} 
 Zeros of $D$ correspond to localized edge states with the energy less than the bottom of the lower band.  For edge states the incident wave  is not present in the wave superposition and  the energy is determined by Eq. (\ref{ener}) with imaginary $k_0=ip$ and positive real $p$. The $x$ components of the wave vector are  complex: $k_\pm =\sqrt{(\alpha \pm  ip)^2-k_y^2}$.  We shall consider the case of small $p$ when the edge states are not deep and one may use the expansion in $p$. The case is realized if $\alpha -k_b \ll \alpha$. Then the edge states with positive $p$ exist in the interval $k_b<|k_y|<\alpha$, where the depth of the edge state (difference between the energies of the band bottom and of the edge state) is
\begin{eqnarray}
\Delta \epsilon ={\hbar^2 p^2\over 2m } = {\hbar^2(|k_y|-k_b)^2(\alpha-|k_y|)\over  \alpha m}~.
         \end{eqnarray} 
The edge states form two degenerate one-dimensional bands, which differ with the sign of the wave number $k_y$. The whole ensemble of these states may be considered as a quantum wire. Because of the crucial role of the spin-orbit interaction for its existence, one may call it {\em spin-orbit quantum wire}. The remarkable feature of the spin-orbit  wire is absence of electrons with small wave numbers $|k_y| <k_b$.

Now we shall demonstrate the appearance  of the spin torque  assuming an infinite barrier at the edge $x=0$ ($k_b \to \infty$). Then the wave function  at $x=0$ vanishes, and  expressions for the  reflection coefficients become simple:
\begin{eqnarray}
r_1={\kappa_+ -\kappa_- \over \kappa_--\bar \kappa_+ }~,~~r_2={ \bar \kappa_+ -  \kappa_+\over  \kappa_--\bar \kappa_+}~.
             \end{eqnarray} 
 Since in a single plane wave the torque vanishes,   only interference of waves can provide a torque. Namely, the $y$ component of the torque originates from interference of the incident wave with the second reflected wave: 
  \begin{eqnarray}
G_{y+}(\vec k)=-{\alpha \hbar^2 k_y\over m} \mbox{Re} \left\{ e^{-ik_+ x +i k_- x}\left(1- \bar \kappa_+  \kappa_- \right)r_2 
\right\}.
     \label{torq}      \end{eqnarray}   
A similar contribution $G_{y-}$ comes from the conjugate superposition,  in which the incident plane wave $\propto e^{-i k_- x}$ corresponds to the wave number $k<\alpha$ with the negative group velocity. Further we restrict ourselves with a simpler case $k_0\ll \alpha$. Then all expression can be expanded in $k_0$, and integrating the single-mode torque over the Fermi sea of the wave vectors $\vec k$ and summing contributions from $k>\alpha$ and $k<\alpha$ one obtains:
\begin{eqnarray}
G_y(x)
= - {4\pi \alpha^2 \hbar^2 k_m^2\over m}\left[\, _1F_2\left(-{1\over 2};1,{3\over 2};-k_m^2 x^2\right)
\right. \nonumber \\ \left.
-{1\over 2} {_1F}_2\left(-{1\over 2};{3\over 2},3;-k_m^2 x^2\right)+{2\over 3}k_mx \right]\, ,
            \end{eqnarray} 
where $k_m$ is the maximum value of $k_0$ corresponding to the Fermi energy and
$_pF_q(a_1,...,a_p; b_1,...,b_q;z)$ is the generalized hypergeometric function \cite{TF}. 
The total torque over the whole bulk   $\int_{-\infty}^0G_y(x)dx=\pi \hbar^2 \alpha^2 k_m/m$   exactly compensates the bulk spin current [see Eq. (19) in Ref. \onlinecite{BR} in the limit $k_m \ll \alpha$].

At large distances from the border the torques for single modes fast oscillate, so the asymptotic behavior of the torque can be analyzed using the steepest-descent method. This yields the asymptotic torque at $x\to -\infty$:
\begin{eqnarray}
G_y =-\sqrt{\pi\over k_m }{\alpha^2\hbar^2\over m |x|^{5/2}}  \sin\left(2k_m x-{\pi\over 4}\right)\, .
             \end{eqnarray}              
The oscillation results from interference between incident and reflected waves. But one should remember that our analysis ignores any disorder or electron-electron interactions, which may suppress oscillations at some distance from the edge.

For numerical estimation one may use the typical value of the spin-orbit coupling $\alpha \hbar^2/m =10^{-9}~\mbox{eV cm}=1.6\cdot 10^{-21}~\mbox{erg cm} $  \cite{R2}. Calculating the spin torque it was simpler to consider the case  $k_m \ll \alpha$. But the  mechanical torque reaches its maximum at $k_m > \alpha$  (see Ref.   \onlinecite{BR}) when the spin current is $J_x^y=-2\pi \alpha ^3 \hbar^2/3m\sim 10^{-8}$  erg/cm.  In order to estimate the displacement $h$ of the cantilever end  (see Fig. \ref{fig1}), we use the cantilever parameters from Ref. \onlinecite{CL}:  the length $l=120~\mu$m and the spring constant $k=F/h=86~\mu$N/m, where $F$ is the force on the cantilever end.  Using the theory of elastic plates \cite{LL}, one obtains that the torque $\tau=-J_x^y$ produces the displacement $h=3 \tau /2 k l$. This yields $h \sim 0.15~\mu$m, which is certainly measurable with the modern micromechanical technique. Moreover, we shall see that the torque can be enhanced and tuned by an external magnetic field. 

In the presence of the external magnetic field the hamiltonian Eq. (\ref{Ham}) must contain the Zeeman energy $-\mu_B \vec \sigma \cdot \vec H$, where $\mu_B=e\hbar/2mc $ is the Bohr magneton. According to Eq. (\ref{SC}), the spin current in a single-electron state is determined by spin in this state, which is parallel or antiparallel to  the ``effective'' magnetic field $\vec H+\alpha \Phi_0 [\vec k \times \hat z]/\pi$ acting on the electron.
Here $\Phi_0=hc/e$ is the single-electron flux quantum.
Integrating the single-state spin current over the whole $\vec k$ space and assuming the Zeeman energy is much larger  than the spin-orbit energy, one obtains the bulk spin current 
\begin{eqnarray}
J^y_x  = \mp \pi \alpha { \epsilon_F ^2-\mu_B^2  H^2\over  \mu_B  H}(1-\sin \theta \sin^2\phi)
 ~,
       \label{h-z}       \end{eqnarray}   
where $\epsilon_F$ is the Fermi energy, $\theta$ is the angle between the $z$ axis and the magnetic field, and $\phi$ is the angle between the in-plane component of the magnetic field and the $x$ axis (direction of the spin current). In the interval $-\mu_B H<\epsilon_F<\mu_B H$  electrons fill only the lower band [the lower sign in Eq. (\ref{h-z})]. Then in terms of the electron density $n=m(\epsilon_F +\mu_BH)/2\pi \hbar^2$
\begin{eqnarray}
J^y_x  =-4 \pi^2\alpha{ \hbar ^2\Phi_0\over m H}n\left( n - {H \over  2\pi \Phi_0}\right) (1-\sin \theta \sin^2\phi)\, .
       \end{eqnarray}   
If $\epsilon_F>\mu_B H$, electrons fill the both bands,  the contributions from two bands to the spin current cancel each other, and spin current vanishes. But this cancellation is valid only in the first order with respect to $\alpha$ and does not rule out spin-currents in higher approximations.

It is worthwhile to comment that ambiguity of spin-current definition, which was intensively discussed  in the literature, has no impact on the effect predicted here. As discussed in Ref. \onlinecite{BR}, one may redefine the spin current $J_j^i$ by adding to it an arbitrary term ($J_j^i \to J_j^i +\delta J_j^i $) but at the same time it is necessary to compensate it by redefinition of the spin torque ($G_i \to G_i +\nabla_j \delta J_j^i $). If the balance of the orbital part of the angular momentum is also considered, the definitions of the orbital torque and flux must be correlated with those in the spin part, in order not to violate the conservation law of the total angular momentum. Eventually  whatever definition used any {\em correct} calculation must predict the same observable effect (displacement of the cantilever).  Our choice of spin current,  Eq.~(\ref{SC}),  coincides with the  definition used by Rashba \cite{R}: $ J^\beta_i ={\hbar \over 4}(\mathbf{\Psi}^\dagger\{ \sigma_\beta  v_i +v_i \sigma_\beta\}\mathbf{\Psi})$,   where $\vec v  ={\hbar\over m}(-i \vec \nabla+ \alpha  [\hat z \times   \vec  \sigma  ]) $
is the operator of the electron group velocity. After this choice we are not free anymore in the choice of the definition of the torque and the current  of the orbital angular momentum: the flux of the the orbital angular momentum in the elastic cantilever should be defined as $\tilde J_j^i =\varepsilon_{imn} x_m T_j^n$ where $T_j^n$ is the elastic stress tensor. This choice looks most natural since it defines the mutual torque between the  spin and the orbital moment as a derivative of the spin-orbit energy with respect to the rotation angle.

In summary, we demonstrated that the equilibrium spin currents  inside the Rashba medium must result in spin torques near edges of the medium.  According to the conservation law for the total angular momentum, the spin torque must lead to the  mechanical torque on the edge of the substrate on which the Rashba medium is formed. If the substrate is a flexible cantilever with one end fixed in a rigid wall, the mechanical torque must displace the cantilever free end. The effect can be enhanced and tuned by applying the magnetic field.
The observation of this displacement would be a direct confirmation of the presence of equilibrium spin currents in the Rashba medium.  The present analysis also predicts possibility of the 1D continuum of the edge states (spin-orbit quantum wire). 

The work was supported by the grant of the Israel Academy of
Sciences and Humanities.

\end{document}